\documentclass[twocolumn,showpacs,preprintnumbers,amsmath,amssymb,APSl,prd,nofootinbib,superscriptaddress]{revtex4-1}

\usepackage{dcolumn}
\usepackage{bm}
\usepackage{ifpdf}
\usepackage{hyperref}
\usepackage{dcolumn}
\usepackage{bm}
\usepackage[spanish,english]{babel}
\usepackage{amsfonts}
\usepackage{amssymb}
\usepackage{graphicx}
\usepackage[latin1]{inputenc}

\newcommand \be{\begin{equation}}
\newcommand \en{\end{equation}}
\newcommand \bea{\begin{eqnarray}}
\newcommand \ena{\end{eqnarray}}

\newcommand \Tr{\mbox{\rm Tr}}

\begin{document}

\title{Crystal clear lessons on the microstructure of space-time and modified gravity}

\author{Francisco S. N.~Lobo} \email{flobo@cii.fc.ul.pt}
\affiliation{Centro de Astronomia e Astrof\'{\i}sica da Universidade de Lisboa, Campo Grande, Edif\'{\i}cio C8, 1749-016 Lisboa, Portugal}
\affiliation{Instituto de Astrof\'{\i}sica e Ci\^{e}ncias do Espa\c{c}o, Universidade de Lisboa, OAL, Tapada da Ajuda, PT1349-018 Lisboa, Portugal.}
\author{Gonzalo J. Olmo} \email{gonzalo.olmo@csic.es}
\affiliation{Departamento de F\'{i}sica Te\'{o}rica and IFIC, Centro Mixto Universidad de
Valencia - CSIC.\\ Universidad de Valencia, Burjassot-46100, Valencia, Spain}
\affiliation{Departamento de F\'isica, Universidade Federal da
Para\'\i ba, 58051-900 Jo\~ao Pessoa, Para\'\i ba, Brazil}
\author{D. Rubiera-Garcia} \email{drubiera@fudan.edu.cn}
\affiliation{Center for Field Theory and Particle Physics and Department of Physics, Fudan University, 220 Handan Road, 200433 Shanghai, China}
\affiliation{Departamento de F\'isica, Universidade Federal da
Para\'\i ba, 58051-900 Jo\~ao Pessoa, Para\'\i ba, Brazil}


\date{\today}

\begin{abstract}
We argue that a microscopic structure for space-time such as that expected in a quantum foam scenario, in which microscopic wormholes and other topological structures should play a relevant role, might lead to an effective metric-affine geometry at larger scales. This idea is supported by the role that microscopic defects play in crystalline structures. With an explicit model we show that wormhole formation is possible in a metric-affine scenario, where the wormhole and the matter fields play a role analogous to that of defects in crystals. Such wormholes also arise in Born-Infeld gravity, which is favoured by an analogy with the estimated mass of a point defect in condensed matter systems. We also point out that in metric-affine geometries Einstein's equations with an effective cosmological constant appear as an attractor in the vacuum limit for a vast family of theories of gravity. This illustrates how lessons from solid state physics can be useful in unveiling the properties of the microcosmos and defining new avenues for modified theories of gravity.
\end{abstract}

\maketitle

\section{Introduction}

Einstein's deep insight on the relation between gravitation and geometry represents the culmination of scientific thinking on the nature of gravitation initiated with the pioneering works of Galileo. The idealization of a body as a point particle moving along geodesics of a space-time metric determined by the matter distributions leads to a number of predictions which are in excellent agreement with observations \cite{Will}. Nonetheless, the approach in terms of point particles necessarily leads to singularities in the metric, which is both conceptually and operationally disturbing. With the notion of {\it geon}, Wheeler \cite{Wheeler} found a way out of this problem. Wheeler's geon is an example of a singularity free classical body consisting on a kind of {\it ball of light}, where the light intensity is so high that the object is held together by its own gravitational field. Geons represent solutions of the Einstein-Maxwell field equations without sources (see \cite{Melvin}). Furthermore, with the seasoning of non-trivial topology, Misner and Wheeler \cite{MW} showed that models of classical particle-like entities with mass and charge could be constructed without resorting to the traditional idea of point-like sources. This inspired the idea that space-time topology could play a fundamental role in the understanding of elementary particles. Combined with the fact that pairs of particles can be spontaneously created out of the quantum vacuum,  one is led to visualize space-time at Planckian scales as a dynamical entity in which all manners of non-trivial topological structures, such as wormholes \cite{Garf-Stro}, are continuously created and annihilated. The space-time continuum that we observe, therefore, could have a highly non-trivial and dynamical microstructure denoted by Wheeler as space-time foam.

However, it is still unknown how to realize such a foam-like space and also whether this represents the real quantum gravitational vacuum. Wheeler when discussing the quantum fluctuations in the space-time metric \cite{Wheeler} considered that a typical fluctuation in a typical gravitational potential is of the order $\Delta g\sim(hG/c^{3})^{1/2}/L$, which becomes appreciable for small length scales $L$. A fundamental question is whether a change in topology may be induced by large metric fluctuations. Indeed, our lack of understanding of this microstructure makes it difficult to visualize how the transition to the smooth continuum that we perceive might take place. It is generally believed that, given the smallness of the Planck length, the details of the microstructure should have little or negligible impact at the scales that we can experimentally probe. A better understanding of the framework necessary for a consistent  transition to macroscales, however, may bring important new insights on the properties of the resulting  dynamics.

In this work, we discuss the geometric elements necessary to describe how this transition from a microstructure to a smooth continuum geometry may occur and its implications for the gravitational dynamics. We pay special attention to the fact that new physics at microscales (high energies) does not necessarily imply the existence of new propagating degrees of freedom. Rather, the approach presented here puts forward that new gravitational physics may arise via non-perturbative effects in such a way that the dynamical equations of General Relativity (GR) become a kind of {\it attractor} to which many different theories of gravity converge at low energies. We provide a variety of examples to illustrate this idea.

\section{Crystalline structures}

Examples of systems with a microscopic structure that admit a smooth continuum limit are well known in solid state physics. Bravais crystals are particularly important for our discussion \cite{Kittel}. The internal crystal structure is composed by a three-dimensional net-like arrangement of atoms (taken as small points), provided that the environment around any of these atoms is the same as around any other. It turns out that the continuized, macroscopic, Bravais crystal can be described using the language of differential geometry \cite{Kroner1,Kroner4}. For ideal and perfect crystals (the latter being simply a deformation of the former), the geometry becomes Riemannian, whereas crystals with defects require a metric-affine approach \cite{Kroner2}. Examples of defects are interfaces (two-dimensional), dislocations/disclinations (one-dimensional), and vacancies/interstitials (point-like). In this framework, dislocations become the discrete version of Cartan's torsion \cite{Kroner3,Kondo,Bilby}, whereas point-like defects involve non-metricity \cite{Falk81}. In fact, a full theory of defects as gauge fields can be established \cite{Kleiner}.

Let us consider, for simplicity, the case with point-like defects. These are of two types: intersticials, which are atoms that have been taken away from their network equilibrium positions, and vacancies, corresponding to the empty places left behind. The continuized crystal is the result of a limiting process in which the lattice spacing and the particle mass are taken to zero while keeping the matter density finite \cite{Kroner1}, and provides a valuable model of the real, discrete crystal. The crystal with defects can also be continuized in this way, provided that the total amount of defects per unit volume is kept unchanged. At each atom of the crystal we can define a triad of vectors (crystallographic directions) such that parallel transport along these directions take us from one atom to another. Inside the crystal, distances can be determined by step counting along the crystallographic directions.

For the ideal crystal, the crystallographic coordinates are Cartesian and have vanishing connection coefficients by definition, $\Gamma^a_{bc}(x)=0$. Since a perfect crystal is a deformation of an ideal one, infinitesimal displacements in the former are related to those of the latter  by $dy^\mu={e^\mu}_i dx^i$, where ${e^\mu}_i$ represents the distorsion of the structure and satisfies ${e^\mu}_i {e^i}_\nu=\delta^\mu_\nu$ and ${e^i}_\mu {e^\mu}_j=\delta^i_j$. Accordingly, the connection coefficients and the line element in curvilinear coordinates become $\Gamma^\alpha_{\beta\gamma}(y)={e^\alpha}_i\partial_\beta {e^i}_\gamma$, and $ds^2=g_{\mu\nu} dy^{\mu} dy^{\nu}$, respectively, where $g_{\mu\nu}\equiv \delta_{ij}{e^i}_{\mu} {e^j}_{\nu}$. In the perfect crystal, therefore, both metric and connection are completely determined by the deformation ${e^\mu}_i$. One can verify that $\nabla^\Gamma_\lambda g_{\mu\nu}=0$, which implies that the connection is metric compatible.

In a crystal with point defects, the step counting procedure for measuring lengths breaks down at the location of the defects. The situation becomes non-metric. To overcome this, one can introduce a pseudo-length measurement with the prescription that one just ignores the point defects, filling vacancies with atoms and not counting interstitials at all \cite{Kroner1}. It can be represented as $d\tilde{s}^2=h_{\mu\nu}dx^\mu dx^\nu$, with $h_{\mu\nu}=\delta_{ij}c^i_{\mu} c^j_{\nu}$ and  $c^i_{\mu}\neq e^i_{\mu}$. A connection $L^\alpha_{\beta\gamma}$ compatible with this metric can be defined. The deformation $d^i_{j}$ that relates the auxiliary structure $c^j_{\mu}$ with the physical one, $e^i_{\mu}=d^i_{j}c^j_{\mu}$, is determined by the kind and  density of defects per unit volume, being $d^i_j=\delta^i_j $ in the case with no defects. It should be noted that point defects have dynamics, being able to move through the crystals upon the action of heat or external fields, and to recombine with another point defect or line defect.

We thus see that the metric $h_{\mu\nu}$ provides a useful notion of distance in the defected crystal and allows to construct $\Gamma^\alpha_{\beta\gamma}(g)$, which contains information about the distribution of defects and serves to define the physical paths in the continuized geometry. Obviously, in a crystal with point defects, we have
$\nabla^\Gamma_\lambda h_{\mu\nu}\neq 0$ (and $\nabla^L_\lambda g_{\mu\nu}\neq 0$). This puts forward that the continuum limit of defected crystals leads to a metric-affine geometry. It should be noted that, though only a neighbourhood around the point defects gets disturbed by their presence, it turns out that the density of defects present in the crystal, which are objects in the microscale, have a profound influence on the emergence and the details of many collective properties in the macroscopic scale of materials. These properties, such as viscosity and viscoelasticity (associated to point defects) or plasticity (for line-like defects) are of great importance for the understanding of materials such as graphene (see \cite{Iorio} for a recent work on interpretation of graphene as a quantum field in curved space-time). In this way, knowledge of the macroscopic properties allows to infer properties from the microscopic structure, and vice-versa.

A fundamental lesson that follows from the above considerations is that the existence of a non-zero density of  point-like topological defects
in the space-time microstructure could lead to a continuum geometry of the  metric-affine type. If the density of defects is not uniform, the geometry may become Riemannian in those regions and/or at those scales where the density of defects is sufficiently low. The success of Einstein's theory of general relativity (GR), which is based on a Riemannian structure, would support the view that the distribution of defects in the space-time microstructure is not uniform and results in an effective Riemannian geometry at the currently accessible scales. Effective descriptions of gravitation at higher energies, however, should take into account the need for going beyond the Riemannian approach. In this sense, we note that metric-affine geometries have been discussed in the literature in the context of torsion and its associated dislocations, while non-metricity has been seldom considered. In this work we intend to modestly fill this gap by establishing a correspondence via non-metricity between  extensions of GR at high energies and the theory of microscopic defects in crystals.

\section{Beyond GR in a non-Riemannian geometry\label{sec:f(R,Q)}}

To delve into the above ideas, and for concreteness, we consider a generic metric-affine extension  of GR of the form
\begin{eqnarray}\label{eq:Quad-Pal}
S[g,L,\psi_m]&=&\frac{1}{2\kappa^2}\int d^4x\sqrt{-g} f(R,R_{\mu\nu}R^{\mu\nu})
   \nonumber \\
  &+&S_m[g_{\mu\nu},\psi_m] \ ,
\end{eqnarray}
where $\kappa^2\equiv 8\pi G/c^4$,
$g_{\mu\nu}$ is the space-time metric, $R=g^{\mu\nu}R_{\mu\nu}$, $R_{\mu\nu}={R^\rho}_{\mu\rho\nu}$, and
${R^\alpha}_{\beta\mu\nu}=\partial_{\mu}
L^{\alpha}_{\nu\beta}-\partial_{\nu}
L^{\alpha}_{\mu\beta}+L^{\alpha}_{\mu\lambda}L^{\lambda}_{\nu\beta}-L^{\alpha}_{\nu\lambda}L^{\lambda}_{\mu\beta} $. The matter action, $S_m=\int d^4x \sqrt{-g} L_m$, is assumed to couple only to the metric $g_{\mu\nu}$, and $\psi_m$ denotes collectively the matter fields. The connection $L^{\alpha}_{\mu\nu}$ is {\it a priori} independent of the metric $g_{\mu\nu}$ (Palatini or metric-affine formalism) and must be determined by the field equations  \cite{OSAT09}
\begin{eqnarray}
f_R R_{\mu\nu}-\frac{f}{2}g_{\mu\nu}+2f_Qg^{\alpha\beta} R_{\mu\alpha}{R}_{\nu\beta} &=& \kappa^2 T_{\mu\nu} \,, \label{eq:met-varX}\\
\nabla^L_{\beta}\left[\sqrt{-g}\left(f_R g^{\mu\nu}+2f_Qg^{\mu\alpha}g^{\nu\beta} R_{\alpha\beta}\right)\right]&=&0  \,,
 \label{eq:con-varX}
\end{eqnarray}
where $T_{\mu \nu}=-\frac{2}{\sqrt{-g}} \frac{\delta (\sqrt{-g} L_m)}{\delta g^{\mu\nu}}$ is the stress-energy tensor of the matter, we have defined $Q\equiv R_{\mu\nu}R^{\mu\nu}$, and used the shorthand notation $f_X\equiv \partial_X f$. For simplicity we have set torsion to zero (see \cite{or-torsion} for a discussion on the inclusion of torsion in these equations), which allows us to focus only on the role of non-metricity. Let us point out that, in general, these field equations are completely different from what one finds in the standard metric (Riemannian) approach \cite{Borunda}, where the connection is taken to be \emph{a priori} given by the Christoffel symbols of the metric.

Now, defining ${P_\mu}^\nu\equiv R_{\mu\alpha}g^{\alpha\nu}$, Eq. (\ref{eq:met-varX}) yields (we denote $\hat{A}\equiv {A_\mu}^\nu $)
\begin{equation} \label{eq:P-quadratic}
2f_Q\left(\hat{P}+\frac{f_R}{4f_Q}\hat{I}\right)^2=\left(\frac{f}{2}+\frac{f_R^2}{8f_Q}\right)\hat{I}+\kappa^2 \hat T \ .
\end{equation}
Given that $R=\Tr[\hat P]$ and $Q=\Tr[\hat P^2]$, this equation implies that $\hat{P}$ is an algebraic function of the stress-energy tensor, i.e., $\hat{P}=\hat{P}(\hat{T})$ .  This allows to turn Eq. (\ref{eq:con-varX}) into $\nabla_{\beta}^L[\sqrt{-h} h^{\mu\nu}]=0$, with $h_{\mu\nu}$ related to  $g_{\mu\nu}$ by a deformation, $g_{\mu\nu}={\Omega_{\mu}}^\alpha h_{\alpha\nu}$, of the form
\begin{equation} \label{eq:h-g}
{\Omega_{\mu}}^\nu=\frac{f_R {\delta_{\mu}}^{\nu} +2f_Q {P_{\mu}}^{\nu}}{\sqrt{\det\left|f_R \hat I +2f_Q \hat P\right|}} \ .
\end{equation}
Since $R$ and $Q$ are functions of $\hat P$ and $\hat{P}=\hat{P}(\hat{T})$, it follows that ${\Omega_{\mu}}^\nu$ is fully determined by ${T_\mu}^\nu$.
Note that the equation $\nabla_{\beta}^L[\sqrt{-h} h^{\mu\nu}]=0$, which follows from variation of the action with respect to $L^{\lambda}_{\mu\nu}$, can be written, after some algebraic transformations, under the more standard form $\nabla_{\beta}^L h^{\mu\nu}=0$. This last equation expresses the compatibility of the connection $L^\alpha_{\beta\gamma}$ with the metric $h_{\mu\nu}$, whereas $\nabla_\alpha^L g_{\mu\nu}\neq 0$, which means that the matter stress-energy density in theories with non-linear curvature corrections generates non-metricity. Using (\ref{eq:met-varX}), one can show that the metric $h_{\mu\nu}$ satisfies a set of Einstein-like equations
\begin{equation}\label{eq:Gmn}
{G_\mu}^\nu(h)=\kappa^2|\hat{\Omega}|^{\frac{1}{2}}\left[{T_\mu}^\nu-\frac{1}{2} {\delta_\mu}^\nu\left(T+2\mathcal{L}_G\right)\right] \ ,
\end{equation}
where $\mathcal{L}_G\equiv f/2\kappa^2$ represents the gravity Lagrangian.

Let us now discuss the vacuum behavior of ${\Omega_{\mu}}^\nu$ and ${G_\mu}^\nu(h)$.  Setting $\hat T=0$ in  Eq. (\ref{eq:P-quadratic}), it follows that $\hat P= C(R,Q)\hat I$, where $C(R,Q)\equiv\sqrt{\left(\frac{f}{2}+\frac{f_R^2}{8f_Q}\right)/2f_Q}-f_R/4f_Q$. Tracing this equation and its square, we get $R=4C(R,Q)$ and $Q=4C^2(R,Q)$. These are two equations with two unknowns that can only be satisfied for particular constant values $R_v$ and $Q_v$, which represent the vacuum configurations. Therefore, $\hat P= C(R_v,Q_v)\hat I$, with $C(R_v,Q_v)$ a constant, which turns Eq. (\ref{eq:h-g}) into
\begin{equation}\label{eq:Om_vac}
{\Omega_{\mu}}^\nu=\frac{1}{\left.\left(f_R+2f_Q C\right)\right|_{v}}{\delta_{\mu}}^{\nu} \ .
\end{equation}
This result shows that in vacuum ${\Omega_{\mu}}^\nu$ is a constant multiple of the identity matrix. As a result, $R_{\mu\nu}(h)=R_{\mu\nu}(g)$ and we find that in vacuum, Eq. (\ref{eq:Gmn}) becomes $G_{\mu\nu}(h)=-\kappa^2|\hat{\Omega}|^{\frac{1}{2}}\mathcal{L}_G h_{\mu\nu}$ or, equivalently,
\begin{equation}\label{eq:Gmn_vac}
 G_{\mu\nu}(g)=-\left.\frac{\kappa^2\mathcal{L}_G}{\left(f_R+2f_Q C\right)}\right|_{v} g_{\mu\nu}=-\Lambda_{\rm eff} g_{\mu\nu} \ ,
\end{equation}
which formally coincide with Einstein's equations in vacuum with an effective cosmological constant $\Lambda_{eff}$.

From the above equations, we find a remarkable formal similarity between the geometric properties of the theory (\ref{eq:Quad-Pal}) and crystalline structures with point-like defects. In the gravitational scenario, the presence of non-linear curvature corrections generates non-metricity (rather than leading to higher-order derivative equations as in the metric formulation). As a consequence, the matter fields build up a GR-like Riemannian structure for $h_{\mu\nu}$ through Eq. (\ref{eq:Gmn}), but the physical metric $g_{\mu\nu}$ is related with $h_{\mu\nu}$ via a deformation  ${\Omega_{\mu}}^\nu$ which depends on the stress-energy densities of the matter fields, see Eq. (\ref{eq:h-g}). In the case of GR, these two metrics are equivalent and one recovers the standard Einstein equations. This is precisely what happens in the vacuum case, where regardless of the form of $\mathcal{L}_G$ the field equations boil down to those of GR$+\Lambda_{eff}$ [see Sec.\ref{sec:IV} below]. The role of the matter stress-energy density, ${T_\mu}^{\nu}$, in these theories is thus akin to the role of defects in crystalline structures. In this sense, the matter fields can be regarded as the continuum, coarse-grained description of microscopic defects in the space-time structure, as they are responsible for the existence of non-metricity ($\nabla_\alpha^L g_{\mu\nu}\neq 0$). To our knowledge, this correspondence has not been established before in the literature and constitutes a relevant motivation for the consideration of theories with non-metricity in gravitational scenarios.

This analogy also extends naturally to vacuum configurations. In the case of crystals, one finds that in regions without defects ($d_a^m=\delta_a^m$) the correspondence  $ {\Omega_{\mu}}^\nu =c^a_\mu d_a^m d_m^b c^\nu_b$  yields ${\Omega_{\mu}}^\nu={\delta_{\mu}}^\nu$ and $g_{\mu\nu}=h_{\mu\nu}$. In the gravitational case, Eq. (\ref{eq:Om_vac}) puts forward that $g_{\mu\nu}$ and  $h_{\mu\nu}$ are related by a constant conformal factor, which can be absorbed into an irrelevant global rescaling of units. The Riemannian condition $\nabla_\alpha g_{\mu\nu}=0$ is thus recovered in vacuum and the notion of metricity as a simple step counting procedure is restored.

\section{Universality of Einstein's equations} \label{sec:IV}

We have just seen that in a crystal without defects parallel transport along crystalographic directions and step counting are compatible operations. The same property, $\nabla_\alpha g_{\mu\nu}=0$, is observed in Palatini $f(R,Q)$ theories in vacuum. Besides this information about whether the underlying geometry is Riemannian or not, the gravitational scenario provides another important piece of information, namely, the dynamics of the resulting geometry. Our discussion of Sec. \ref{sec:f(R,Q)} started with the gravitational theory (\ref{eq:Quad-Pal}), characterized by an unspecified Lagrangian $f(R,R_{\mu\nu}R^{\mu\nu})$, and ended with a generic vacuum dynamics of the form (\ref{eq:Gmn_vac}), which coincides with the dynamics of GR in the presence of an effective cosmological constant $\Lambda_{\rm eff}$. It is important to note that these equations are in sharp contrast with the dynamics corresponding to the traditional metric formalism (where the connection is taken to be compatible with the metric \emph{a priori}) for Lagrangians of the form $f(R,R_{\mu\nu}R^{\mu\nu})$, where higher-order derivative equations appear and their properties depend on the specific Lagrangian chosen \cite{metricapproach}. In the Palatini case, regardless of the form of the function $f(R,R_{\mu\nu}R^{\mu\nu})$, the vacuum dynamics is always captured by Einstein's equations. Interestingly, this is not a unique property of this family of theories. In fact, if one considers a gravitational Born-Infeld (BI) Lagrangian {\it \`{a} la} Palatini \cite{Deser,Banados}, which is not of the  form $f(R,R_{\mu\nu}R^{\mu\nu})$, the vacuum equations also recover GR+$\Lambda_{\rm eff}$. If one adds to the BI Lagrangian an $f(R)$ term, the resulting vacuum equations also recover  GR+$\Lambda_{\rm eff}$ \cite{moo}. Considering $f(|\hat \Omega|)$ deformations of the BI Lagrangian, being the BI theory the case $f(|\hat \Omega|)=|\hat \Omega|^{1/2}$, in vacuum one also gets GR+$\Lambda_{\rm eff}$ \cite{oor}. In another completely different family of Palatini theories inspired by massive gravity and BI gravity, the vacuum result is also GR+$\Lambda_{\rm eff}$ \cite{bho}.

The recovery of Einstein's equations in vacuum, therefore, seems to be a rather generic property of Palatini theories \cite{Mauro}. This observation is relevant for several reasons. In gravitational scenarios, it relaxes the need for a strictly constant $\Lambda_{\rm eff}$, being now a function that in the vacuum limit tends to a specific constant value. The low density of the Universe in our current expansion phase would thus justify the existence of an effective cosmological constant term. The fact that $\Lambda_{\rm eff}$ may arise as the evolution of a function in the vacuum limit and needs not be strictly constant, as in the metric formulation of GR, may help reconsider the traditional interpretation of the cosmological constant as a manifestation of quantum vacuum energy. Note, in this sense, that requiring second-order equations and covariance in the traditional metric formalism leads to GR+$\Lambda$, with $\Lambda$ strictly constant, as the unique solution in four dimensions. In a metric-affine scenario, however, the same dynamical equations appear generically in the low density limit without requiring the constancy of the term that produces $\Lambda_{\rm eff}$.

On the other hand, the stress and strain tensors in the theory of elasticity in solids can be put into correspondence with the Einstein tensor and its conservation laws in three dimensions \cite{Kroner2,Katanaev:1992kh}. The relevance of Einstein's equations (and their extension to the Einstein-Cartan case when line defects such as dislocations and disclinations are considered) in crystalline structures is thus manifest. This suggests that Einstein's equations play a very fundamental role in spaces with independent metric and affine structures.

\section{Wormholes as point defects}

In a quadratic theory of the form (\ref{eq:Quad-Pal}), where $f=R+l_P^2(a R^2+bR_{\mu\nu}R^{\mu\nu})$ and $l_P \equiv \sqrt{\frac{\hbar G}{c^3}}$ represents the Planck length, coupled to a sourceless electric field with Lagrangian $\mathcal{L}_m=-\frac{1}{2} F_{\mu\nu}F^{\mu\nu}$,  it was found \cite{or} that the point-like central singularity of the electrovacuum  solutions of GR (Reissner-Nordstr\"{o}m black holes) is replaced by an extended structure of area $A=4\pi r^2_c$, where  $r_c=\sqrt{l_P r_q}$ and $r_q^2=2Gq^2$ represents a length scale associated to the electric charge of the solution. The area of the 2-spheres in this space-time can be written as $A=4\pi r^2(x)$, where
\begin{equation}
r^2(x)=\frac{x^2+ \sqrt{x^4+4r_c^4}}{2} \ .
\end{equation}
Clearly, the area attains a minimum at $x=0$. For specific values of the charge-to-mass ratio, one finds that the geometry at $x=0$ is completely smooth, which naturally motivates the extension of the coordinate $x$ to the whole real axis, i.e., $x \in (-\infty, +\infty)$. As a result, a wormhole structure arises, with the throat located at $x=0$. Further exploration of these solutions puts forward that  the electric flux at $x=0$ possesses universal (topological) properties, which allows to conclude that the wormhole (topological) structure exists for arbitrary values of the charge and mass parameters. This is so despite the fact that curvature scalars constructed with the metric $g_{\mu\nu}$ typically diverge at $x=0$. The absence of point-like sources in this problem, combined with the existence of an electric flux trapped in the non-trivial topology of the wormhole, confirms that these solutions are geons in Wheeler's sense. Using a simple model of charge collapse, it was shown in \cite{dynamical} that these wormholes can also be generated dynamically out of Minkowski space, thus supporting the idea of topology change on which the notion of space-time foam is based.

Identical geonic solutions as in the quadratic gravity theory just mentioned also appear in Palatini Born-Infeld gravity theory  \cite{Deser,Banados}, as shown in \cite{ors}. The action of this theory can be written as
\begin{equation}\label{eq:BI}
S_{BI}=\frac{1}{2\kappa^2l^2_\epsilon}\int d^4x \left(\sqrt{-g}- \sqrt{-q}\right)+S_m \ ,
\end{equation}
where $q$ denotes the determinant of a rank-two tensor $q_{\mu\nu}=g_{\mu\nu}-2l^2_\epsilon R_{\mu\nu}$, and $l_\epsilon$ is a length scale, assumed to be small. In this theory, one finds that the connection equation takes the form $\nabla_\alpha^\Gamma\left[\sqrt{-q}q^{\mu\nu}\right]=0$. As a result,  $\Gamma^\alpha_{\mu\nu}$ turns out to be the Levi-Civita connection of $q_{\mu\nu}$, which now plays the role of auxiliary metric. Using the metric field equations, the relation between  $q_{\mu\nu}$ and  $g_{\mu\nu}$ can be written as  $q_{\mu\nu}={{\tilde{\Omega}}_\mu}^\alpha g_{\alpha\nu}$, with ${{\tilde{\Omega}}_\mu}^\alpha$ being a function of the matter fields (see \cite{ors,oor} for details). Similarly as in $ f(R,R_{\mu\nu}R^{\mu\nu})$ theories, here the matter is also responsible for the existence of non-metricity, and only in vacuum does one find $q_{\mu\nu}=g_{\mu\nu}$ (a conformal factor associated to the cosmological constant has been omitted for simplicity).

The Born-Infeld theory in the representation (\ref{eq:BI}) allows to establish one further (new) relation with the theory of point defects in condensed matter systems. It turns out that the mass associated to a point defect is proportional to its volume. A way to measure it is just to compare the volume defined by a defected structure with the volume expected in the perfect or ideal case (see \cite{Katanaev:1992kh} for a discussion in three dimensional structures). The gravitational part of the Born-Infeld action (\ref{eq:BI}) is essentially capturing this same idea, namely, a comparison of invariant volume elements. In the vacuum theory, analogous to the undefected crystal, we find that $\sqrt{-g}- \sqrt{-q}=0$, which simply confirms that there are no point defects. When the electric field is turned on (by adding the matter Lagrangian), we find that $\sqrt{-g}- \sqrt{-q}=\sqrt{-g}(1-{|\tilde\Omega|^{1/2}})$, thus implying that the presence of matter fields is causing a mismatch between the volume elements. In the limit $l_\epsilon\to 0$, the leading order term of $(1-{|\tilde\Omega|^{1/2}})$ reduces to the Ricci scalar, putting forward that the GR action is just considering the zeroth order effects of microscopic defects on the space-time dynamics.

Denoting the action (\ref{eq:BI}) as $S_{BI}=\int dt \int d^3x \mathcal{E}_{BI}$, where $\mathcal{E}_{BI}$ includes both the gravitational and the electromagnetic Maxwell Lagrangian densities, one finds \cite{ors}
\begin{equation}\label{eq:mass}
\int d^3x \mathcal{E}_{BI}=2Mc^2\frac{\delta_1}{\delta_c} \ ,
\end{equation}
where $\delta_1\equiv \frac{1}{2r_S}\sqrt{r_c^3/l_\epsilon}$ specifies the charge-to-mass ratio, where $r_S=2M$ is Schwarzschild radius, and $\delta_c\equiv \frac{3\Gamma[3/4]^2}{\sqrt{2}\pi^{3/2}}\approx 0.572 $. Remarkably this result puts forward that the action (\ref{eq:BI}) evaluated on the solutions recovers the action of a point-like particle of mass $2M\delta_1/\delta_1^*$ at rest.
The choice $\delta_1=\delta_c$ corresponds to the regular solutions mentioned earlier, whereas $\delta_1\neq \delta_c$ represents the solutions where curvature scalars diverge at the wormhole throat. From Eq. (\ref{eq:mass}) it follows that the effective energy of these objects is always finite, regardless of the existence or not of curvature divergences at the throat. The fact that in a simplified electromagnetic setting we have found the presence of wormhole solutions  with properties of point-like particles, further strengthens the suitability of the metric-affine framework to deal with point defects. The correspondence established here between the Born-Infeld gravity action and the mass of a point defect provides an additional motivation to seek in condensed matter systems new sources of inspiration to deal with gravitational questions.

We point out that the size of the wormhole is measured in units of $l_\epsilon$, which somehow characterizes the microscopic space-time structure (like the lattice spacing in crystals). In fact, one can verify that $r_q$ can be written as $r_q=2l_P N_q/N_c$, where $N_q$ represents the number of charges (in proton units) and $N_c\equiv \sqrt{2/\alpha_{em}}\approx 16.55$ is a constant associated to the fine structure constant $\alpha_{em}$. We thus find that $r_c=\sqrt{r_q l_\epsilon}\propto N_q^{1/2}(l_P l_\epsilon)^{1/2}$. If $l_\epsilon$ is identified with the Planck length, then $r_c  \propto N_q^{1/2}l_P$. The wormhole, therefore represents a hole in the space-time structure whose size is measured in the natural units of an analogous microscopic lattice. The intuitive identification with a point-like defect in a condensed matter system is thus manifest. The electric field that generates and threads the wormhole, therefore, has a direct impact on the microscopic structure, but also defines a density of defects over the whole space-time by means of the non-metricity associated to its energy density, as captured by the continuum average that the stress-energy tensor represents. In other words, it has local and global properties.

The fact that these wormholes are generated by an electric field has additional far reaching implications. It turns out that for observers on one side of the wormhole mouth the electric field appears as generated by positive charges, while for observers on the other side it seems to be negatively charged. Therefore, the wormhole is representing at the same time a pair of opposite charges. Note, in this sense, that the factor $2$ appearing on the right-hand side of  Eq. (\ref{eq:mass}) is due to the fact that the wormhole has two identical sides.  The possibility of describing two apparently different objects (positive and negative charges, or intersticials and vacancies) with a single structure (wormhole) has non-trivial implications for the understanding of quantum entanglement \cite{Lobo:2014fma}. The apparent non-locality of entanglement transfer can be interpreted as the fact that acting on one of the manifestations of the wormhole necessarily has an instantaneous effect on the other manifestations of the object. The relation between wormhole geometries and entanglement was first suggested in \cite{Maldacena:2013xja}.

\section{Summary and conclusions}

In this work we have provided an empirical motivation for the consideration of metric-affine theories of gravity with non-metricity in gravitational scenarios. The fact that such geometries arise as emergent phenomena in systems with a microstructure has allowed us to interpret recent results involving high-energy physics in gravitational scenarios  in terms of concepts proper of condensed matter systems. In this way, we have found evidence that relates wormholes with point-like defects in the microstructure of space-time. The Born-Infeld gravity Lagrangian has been particularly useful to establish this relation, as the spatial integral of the total action provides a measure of the mass/energy of the associated solution, in clear correspondence with formulae used in condensed matter systems to estimate the mass of point defects \cite{Katanaev:1992kh}.

The need for an auxiliary metric structure to properly define parallel transport  in crystals with point defects finds a direct correspondence with the existence of an auxiliary metric associated with the connection in Palatini theories of gravity. In both cases the physical and auxiliary metrics are related by a deformation, which in the former case depends on the density of defects, while in the latter depends on the stress-energy tensor of the matter fields. This relation motivates the identification of the matter fields as the continuum coarse-grained description of microscopic defects in a hypothetical microstructure of space-time.

We have pointed out that in vacuum configurations the metric and affine structures are fully determined by the metric structure, i.e., the metricity (Riemannian) condition $\nabla_\alpha g_{\mu\nu}=0$ is recovered.  This is so because in the vacuum case the deformation relating the physical and auxiliary metrics boils down to the identity (up to an irrelevant constant conformal factor, which just amounts to a global redefinition of units). The same occurs in undefected crystals, where the step-counting procedure is well defined everywhere and there is no need for an auxiliary metric structure. Interestingly, the gravitational dynamics in the vacuum case for a vast family of metric-affine theories of gravity coincides with the field equations of GR (with possibly a cosmological constant, depending on the particular Lagrangian). These two facts, namely, the recovery of a Riemannian structure and the appearance of the GR+$\Lambda_{\rm eff}$ equations in vacuum are different and logically independent properties of these theories. Though the reasons for the emergence of the GR+$\Lambda_{\rm eff}$ equations is not well understood, it suggests that Einstein's theory might not be as unique as usually claimed within the traditional metric formalism. Rather, these equations seem to be an attractor to which many different theories converge in the low density or vacuum limit. This is in sharp contrast with the widespread idea that extensions of GR necessarily introduce new dynamical degrees of freedom.

To conclude, we would like to note that the notions of curved space and Riemannian geometry made their first appearance in Physics through Einstein's theory of gravity. However,  different types of effective geometries are now known to play an important role in condensed matter systems, being the Riemannian case associated to idealized configurations without structural defects. Lessons from crystalline structures, which require non-metricity and torsion for their proper description in relation to the existence of defects, could thus serve as useful guides to address the problems that Einstein's theory is facing nowadays.

\section*{Acknowledgments}

 FSNL acknowledges financial support of the Funda\c{c}\~{a}o para a Ci\^{e}ncia e Tecnologia through an Investigador FCT Research contract, with reference IF/00859/2012, funded by FCT/MCTES (Portugal), and the grant EXPL/FIS-AST/1608/2013. G.J.O. is supported by the Spanish grant FIS2011-29813-C02-02, the Consolider Program CPANPHY-1205388, the JAE-doc program of the Spanish Research Council (CSIC), and the i-LINK0780 grant of CSIC. D.R.-G. is supported by the NSFC (Chinese agency) grant No. 11305038, the Shanghai Municipal Education Commission grant for Innovative Programs No. 14ZZ001, the Thousand Young Talents Program, and Fudan University. G.J.O and D.R.-G. also acknowledge funding support of CNPq project No. 301137/2014-5.

\end{document}